\documentclass[aps,prl,reprint,preprintnumbers,showpacs,showkeys,superscriptaddress]{revtex4-1}
\usepackage{amsmath,amssymb,bm,mathrsfs,graphicx}
\usepackage[colorlinks=true,citecolor=blue,linkcolor=blue]{hyperref}

\begin{document}
\title{Criterion for stability of Goldstone Modes and Fermi Liquid behavior in a metal with broken symmetry}
\author{Haruki Watanabe}
\affiliation{Department of Physics, University of California,
  Berkeley, California 94720, USA}
\author{Ashvin Vishwanath}
\affiliation{Department of Physics, University of California, Berkeley, CA 94720, USA}
\affiliation{Materials Science Division, Lawrence Berkeley National Laboratories, Berkeley, CA 94720}

\begin{abstract}
There are few general physical principles that protect the low energy excitations of a quantum phase.  Of these,  Goldstone's theorem and Landau Fermi liquid theory are the most relevant to solids. We investigate the stability of the resulting gapless excitations - Nambu Goldstone bosons (NGBs) and Landau quasiparticles - when coupled to one another, which is of direct relevance to metals with a broken continuous symmetry. Typically, the coupling between NGBs and Landau quasiparticles vanishes at low energies leaving the gapless modes unaffected. If however the low energy coupling is non-vanishing, non-Fermi liquid behavior and overdamped bosons are expected. Here we prove a general criterion which specifies when the coupling is non-vanishing. It is satisfied by the case of a nematic Fermi fluid,  consistent with earlier  microscopic calculations. In addition, the criterion identifies a new kind of symmetry breaking - of magnetic translations - where non-vanishing couplings should arise, opening  a new route to realizing non-Fermi liquid phases. 
\end{abstract}
\maketitle

According to the Goldstone theorem, spontaneous breaking of a continuous symmetry leads to gapless Nambu-Goldstone bosons (NGBs). In a Lorentz invariant theory, these bosons are expected to be well defined excitations, even in the presence of other gapless fields, such as massless Dirac fermions, providing a powerful general mechanism for low energy excitation~\cite{Weinberg}. A key ingredient leading to their stability is the fact that interactions with NGBs are strongly constrained by symmetry, leading to suppressed couplings at small momentum transfer.

In non-relativistic systems though, such general results are not applicable. A particularly important scenario is spontaneous symmetry breaking in a metallic environment, of which there are numerous examples such as magnetic order in a  metal. Do the NGBs then survive as well defined low energy modes? Or does coupling to the high density of gapless fermionic excitations of the metal lead to overdamped excitations? In this work we will establish a general criterion to answer this question based on the pattern of symmetry breaking.

A closely related question has to do with the stability of the Fermi liquid (FL) when coupled to gapless bosonic modes. 
Besides NGBs, gauge bosons can be gapless over an entire phase, i.e. photons of the electromagnetic field or emergent gauge bosons  of spin liquids or quantum Hall states. Alternately, one can tune to a quantum critical point where gapless critical modes centered at wave vector $q=0$ interact with the FL. The latter two cases, of gauge bosons or $q=0$ quantum critical bosons coupled to a Fermi sea have been studied in many works~\cite{HLR,Stern:1995p1,AltshulerIoffeMillis,NayakWilczekNFL,NayakWilczekNFL2,Chakravarty1995,Motrunich,SSLee,SSLee2,RechPepinChubukov, Sachdev1,Sachdev2,Senthil, RaghuNFL1,RaghuNFL2,Fradkin}.  These studies conclude that, for example in $d=2+1$ dimensions the lifetime of excitations near the Fermi surface is significantly reduced, leading to an absence of well defined quasiparticles and a breakdown of FL theory.  Similarly, the bosonic modes get overdamped and can no longer be observed as well-defined particle-like excitations. In some cases however, superconductivity intervenes at low energies~\cite{Metlitski2014}.

In contrast, coupling electrons in a metal to NGBs typically leads to a much more benign outcome. We know from examples of magnons in ferromagnets and phonons in crystals, that NGBs are typically underdamped even in a metallic environment, and the FL theory remains valid.  In other words, in these cases the coupling between NGBs and FLs  is irrelevant, leading to effectively independent fermionic excitations and NGBs at low energies. This is because interactions involving NGBs are very strongly restricted by both broken and unbroken symmetries. In particular, for these cases the scattering amplitude of electrons off NGBs in the limit of small energy-momentum transfer must vanish.  
In contrast, quantum critical modes and gauge bosons couple directly to fermions, without the derivative coupling.

However, there is one known exception to this rule.  When the continuous spatial rotation in $d=2+1$ dimensions is spontaneously broken by a Fermi surface distortion~\cite{Oganesyan2001,Lawler2006,Cenke}, the resulting orientational NGB strongly couples to electrons; {\it i.e.}, their coupling does {\em not} vanish in the limit of small energy-momentum transfer.  We refer to this type of couplings as {\it nonvanishing couplings}.  In this context, Oganesyan, Fradkin and Kivelson~\cite{Oganesyan2001} discussed non-Fermi liquid (NFL) behavior and Landau damping of the NGBs, in close analogy with the case of critical bosons or gauge bosons coupled to a FL. However, the deeper reason why this example violates the standard rule of vanishing NGB-electron couplings in the infrared, has been left unclear. Also, whether this is the only pattern of symmetry breaking with nonvanishing coupling, remains an open question.
  
In this Letter we formulate a simple criterion that allows one to diagnose the nature of the NGB-electron coupling. If the {\em broken symmetry generator fails to commute with translations}, the coupling is anomalous and is nonvanishing in the infrared. Furthermore, armed with this criterion we are able to identify a new physical setting, distinct from the spontaneous breaking of rotation symmetry, that also leads to nonvanishing couplings, and thus, if we follow standard arguments, to a NFL and overdamped NGBs.

\section{The General Criterion for Nonvanishing Couplings}
Let us assume that we are at zero temperature and we make no assumption about spatial dimensionality except that it allows for spontaneous symmetry breaking. NGBs can be associated with symmetry generators that are spontaneously broken, which we label $Q_a$. Furthermore, to sharply define a Fermi surface we assume the existence of  a conserved momentum $\vec{P}$. This could be either the conserved momentum of continuous translation symmetry, or crystal momentum (of discrete translation symmetry).   Let
\begin{equation}
[Q_a,P_i]=i\Lambda_{ai}.
\label{commutator}
\end{equation}
We now state the general criterion. 
If $\Lambda_{ai}=0$, this is the usual situation where the coupling {\em does} vanish. However,  if $\Lambda_{ai}\neq0$ then the coupling between the NGB and electrons does {\em not} vanish in the limit of small energy-momentum transfer. Note, this criterion is very general and only involves the pattern of symmetry breaking. For any internal symmetry (for example spin rotation or number conservation), the commutator is zero. Thus, for nonvanishing couplings one must consider a space dependent symmetry. The simple case of broken space translation symmetry has $Q_a=P_a$ and the commutator is again zero, which implies that the corresponding Goldstone modes, the phonons, have vanishing coupling to electrons at small momentum transfer, as is well known.
 
However, for the case of rotational symmetry breaking, $Q_a = L_z$, which satisfies $[L_z,P_i]=i\epsilon_{ij}P_j\neq0$ where $i,\,j\in\{x,\,y\}$.  Thus, nonvanishing couplings are expected in this case, consistent with the results of Oganesyan-Fradkin-Kivelson in the context of nematic order in a 2D Fermi fluid~\cite{Oganesyan2001,Lawler2006,Cenke}.

\begin{figure}
\begin{center}
\includegraphics[width=\columnwidth,clip]{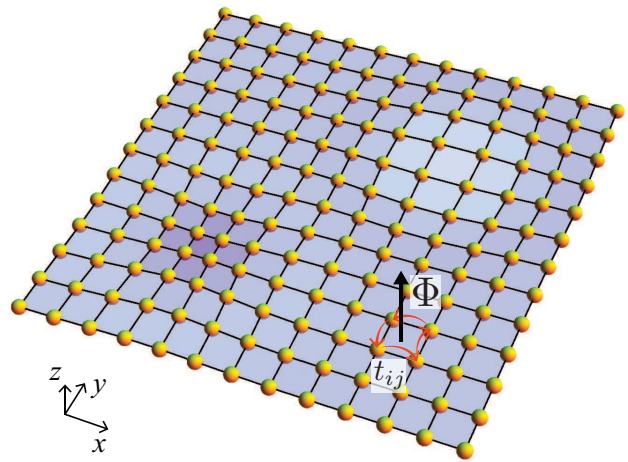}
\end{center}
\caption{A new route to strongly coupling NGBs and quasiparticles - electron-phonon interaction in a uniform magnetic field. A lattice distortion (phonon fluctuation) with $\vec{\nabla}\cdot\vec{u}\neq0$ changes the local flux threading each unit cell (darker blue indicates a larger flux), inducing fluctuations of the phase of the hopping matrix $t_{ij}$. The phonon therefore couples like a gauge field to the fermions, without  spatial derivatives.}
\label{fig:lattice}
\end{figure}

This general criterion allows us to identify an entirely new example of nonvanishing coupling. The criterion for nonvanishing coupling is also fulfilled by spontaneous breaking of {\em magnetic} translations. That is, begin with charged particles in a uniform magnetic field, with magnetic translation symmetry. Spontaneous formation of a crystal breaks this symmetry, resulting in phonons. Now, the magnetic translation operator $\vec{\mathcal P}$ generates NGBs (phonons) and satisfies the non-abelian algebra, $[{\mathcal P}_x,{\mathcal P}_y]=-ieBN$.   Thus electron-phonon interactions under a uniform magnetic field are predicted to have nonvanishing coupling as we verify by explicit calculation. This surprising conclusion may be rationalized by imagining the fermions to hop between sites of the corresponding tight-binding model. The external magnetic  field affects the phase of the hopping matrix as $t_{ij}\exp\big(i\int_{\vec{x}_i}^{\vec{x}_j}\vec{A}(\vec{x}',t)\cdot\mathrm{d}\vec{x}'\big)$.  However, a phonon fluctuation $\vec{u}$ which changes the local flux per a unit cell and produces a fluctuation of $t_{ij}$, as illustrated in Fig.~\ref{fig:lattice}.  One can imaging this as resulting from a fluctuating gauge field $\Delta B = B\nabla\cdot \vec{u}=\nabla\times \delta A$. Therefore, for electrons, some part of phonon fluctuation under a magnetic field is equivalent to that of a vector potential $\delta A = B\hat{z}\times \vec{u}$ and the problem resembles that of NFL behavior arising from minimal coupling to a fluctuating gauge field.

\section{Proof of the General criterion}
The total Hamiltonian of the system can be split into three pieces, $\mathcal{H}_{\text{tot}}=\mathcal{H}_{\text{el}}+\mathcal{H}_{\text{NGB}}+\mathcal{H}_{\text{int}}$, and each of these terms commutes with symmetry generators.  We will mainly be concerned with $\mathcal{H}_{\text{int}}$ which we expand as a series in the NGB fields $\pi^a$, $\mathcal{H}_{\text{int}}=\mathcal{H}_{\text{int}}^{(0)}+\mathcal{H}_{\text{int}}^{(1)}+\cdots$.  Note, $\mathcal{H}_0\equiv\mathcal{H}_{\text{el}}(\bar{\psi},\psi)+\mathcal{H}_{\text{int}}^{(0)}(\bar{\psi},\psi)$ is the mean field Hamiltonian, that defines a one electron problem by picking a symmetry broken ground state. The interaction with NGBs is then determined by symmetry, e.g. the linear coupling for a constant $\pi^a$ is simply obtained by rotating the mean field Hamiltonian by the corresponding symmetry generator $Q_a$ (see~\cite{SM} for details). 
\begin{equation}
\mathcal{H}_{\text{int}}^{(1)}= -[i\pi^aQ_a,\mathcal{H}_0].\label{commutation}
\end{equation}

\begin{figure}
\begin{center}
\includegraphics[width=0.7\columnwidth,clip]{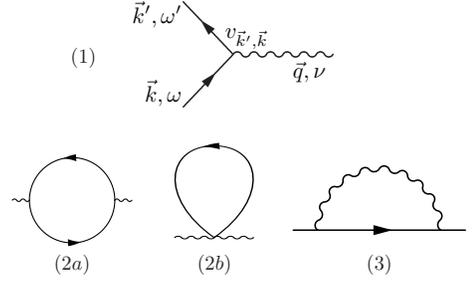}
\end{center}
\caption{(1): The bare vertex with one NG line.  (2a) and (2b): 1-loop diagrams for the self-energy of boson. (3): the same for electrons.}
\label{fig:vertex}
\end{figure}

To setup the perturbation theory, we first solve the single-particle electron problem described by $\mathcal{H}_0\equiv\mathcal{H}_{\text{el}}(\bar{\psi},\psi)+\mathcal{H}_{\text{int}}^{(0)}(\bar{\psi},\psi)$ and obtain simultaneous eigenstates $|n\vec{k}\rangle$ of $\mathcal{H}_0$ and the momentum $\vec{P}$,
\begin{eqnarray}
\mathcal{H}_0|n\vec{k}\rangle=\epsilon_{n\vec{k}}|n\vec{k}\rangle,\quad \vec{P}|n\vec{k}\rangle=\vec{k}|n\vec{k}\rangle,
\end{eqnarray}
where $n$ is the band index. When the translation symmetry is discrete, we replace the second relation with
\begin{eqnarray}
T_i|n\vec{k}\rangle=e^{i\vec{k}\cdot\vec{a}_i}|n\vec{k}\rangle,\quad T_i=e^{i\vec{P}\cdot\vec{a}_i}.
\end{eqnarray}
where $\{\vec{a}_i\}_{i=1,\ldots,d}$ are primitive lattice vectors.  The interaction of electrons and NGBs to lowest order can then be written as (see Fig.~\ref{fig:vertex} (1)):
\begin{equation}
\mathcal{H}_{\text{int}}^{(1)} = \sum_{n',n,a}\int\frac{\mathrm{d}^dk\mathrm{d}^dk'}{(2\pi)^{2d}}v_{n'\vec{k}',n\vec{k}}^ac_{n'\vec{k}'}^\dagger c_{n\vec{k}}\pi_{\vec{q}}^a,
\end{equation}
where $\vec{q}=\vec{k}-\vec{k}'$ and  $v_{n'\vec{k}',n\vec{k}}^a$ is the (bare) vertex function, which is the matrix element of $\mathcal{H}_{\text{int}}^{(1)}$. This can be written via Eqn.~\ref{commutation} as:
\begin{eqnarray}
\pi_{\vec{q}}^av_{n'\vec{k}',n\vec{k}}^a&=&-i\pi^a_{\vec{q}}\langle n'\vec{k}'|[Q_a,\mathcal{H}_0]|n\vec{k}\rangle.\nonumber\\
&=&i\pi^a_{\vec{q}}\langle n'\vec{k}'|Q_a|n\vec{k}\rangle(\epsilon_{n'\vec{k}'}-\epsilon_{n\vec{k}}),
\label{VME}
\end{eqnarray}
which, for low energy scattering: $n=n'$ and $\vec{q}\rightarrow 0$ is 
\begin{equation}
v_{n\vec{k}',n\vec{k}}^a \approx i\langle n\vec{k}|Q_a|n\vec{k}\rangle \;\vec{q}\cdot \vec{\nabla}_{\vec{k}}\epsilon_{n\vec{k}}.
\label{vanishing}
\end{equation}
Clearly, as long as $\langle n\vec{k}|Q_a|n\vec{k}\rangle$ is {\em finite}, the vertex vanishes as $\vec{q}\rightarrow 0$.  This is why scatterings of electrons off NGBs usually vanish at $\vec{q}=0$, leaving behind well-defined NGBs and Fermi liquid quasiparticles. 

However, we can evade this conclusion if, and only if the matrix element $\langle n\vec{k}'|Q_a|n\vec{k}\rangle$ {\em diverges} as $\vec{k}'\rightarrow \vec{k}$. To ensure an appropriate divergence, we will need $[Q_a,P_i]\neq0$ as we now explain. If  $[Q_a,P_i]=i\Lambda_{ai}\neq0$, then the matrix element: $\langle n\vec{k}'|Q_a|n\vec{k}\rangle =-i\frac{\langle n\vec{k}'|\Lambda_{ai}|n\vec{k}\rangle}{k_i'-k_i}$. Substituting this in Eqn.~\ref{vanishing}, and setting $\vec{q}\rightarrow 0$ we have the coupling:
\begin{equation}
v_{n\vec{k},n\vec{k}}^a = \sum_i\langle n\vec{k}|\Lambda_{ai}|n\vec{k}\rangle \partial_{k_i}\epsilon_{n\vec{k}},\label{vertexkk}
\end{equation} 
which is generically nonvanishing. This proves our claim that when the symmetry generator corresponding to the NGBs fails to commute with translations, nonvanishing couplings result. If the translation symmetry is discrete rather than continuous, we simply replace $P^i$ by the discrete translation operator $T_i$, and require $[Q_a,T_i]\neq0$. Then the matrix element 
\begin{eqnarray}
\langle n\vec{k}'|Q_a|n\vec{k}\rangle = -\frac{\langle n\vec{k}'|[Q_a,T_i]|n\vec{k}\rangle}{e^{i\vec{k}'\cdot\vec{a}_i}-e^{i\vec{k}\cdot\vec{a}_i}}\label{vertexdiscrete}
\end{eqnarray}
is inversely proportional to $(\vec{k}'-\vec{k})\cdot\vec{a}_i$ leading again to a non vanishing coupling in Eqn.~\ref{vanishing} as $\vec{k}'\rightarrow \vec{k}$. Note that generically $[Q_a,T_i]\neq0$ follows from $[Q_a,P_i]\neq0$.  Before turning to specific examples we discuss consequences of the non vanishing couplings. 

\section{Non-Fermi Liquid and overdamped Nambu-Goldstone Bosons}
A non-vanishing coupling connects our problem to well-studied problems of a Fermi surface interacting with gauge or critical bosons.  The vertex $\vec{v}_{\vec{k}',\vec{k}}=-e(\vec{k}'+\vec{k})/2m$ of the gauge coupling $-e\vec{A}\cdot\vec{j}$ does not vanish at $\vec{k}'=\vec{k}$.  Similarly, the Yukawa interaction between $q=0$ critical bosons and electrons are not severely restricted by symmetries and nonvanishing couplings are expected (\text{e.g.}, Yukawa couplings).  We can readily argue, via the one-loop calculation below, that nonvanishing couplings destabilize the fixed point of free NGBs and a decoupled Fermi liquid. The actual fate of this strongly coupled problem - a non-Fermi liquid, a superconductor or some other state - requires a case by case analysis and is currently under active investigation.

 The boson self-energy correction $\Pi_{ab}(\nu,\vec{q})$ from the diagrams (2a) and (2b) of Fig.~\ref{fig:vertex} is dominated by~\cite{SM}
\begin{eqnarray}
\Pi_{ab}(\nu,\vec{q})&=&-i\pi\frac{\nu}{|\vec{q}|}\gamma^{ab}(\hat{q}) \label{oneloop1}\\
\gamma^{ab}(\hat{q})&=&\int\frac{\mathrm{d}^dk}{(2\pi)^d}v_{\vec{k},\vec{k}}^av_{\vec{k},\vec{k}}^b\delta(\epsilon_{\vec{k}})\delta(\hat{q}\cdot\vec{\nabla}_{\vec{k}}\epsilon_{\vec{k}}).\nonumber
\end{eqnarray}
The first delta function puts the electron momentum $\vec{k}$ on the Fermi surface and the second one further restricts $\vec{k}$ into a subspace where $\vec{q}$ is tangential to the Fermi surface.  Note that the correction in Eqn.~\ref{oneloop1} vanishes if $v_{\vec{k},\vec{k}}^a=0$ (we have suppressed the band index $n$). The 1-loop corrected boson propagator $D^{-1}=D_0^{-1}-\Pi$ has over-damped poles $\nu\propto-iq^3$ due to the singularity in Eqn.~\ref{oneloop1}. Thus the NGBs are destroyed (overdamped) by interaction with the Fermi surface at this order. 

Now, one can study the lifetime of fermionic quasiparticles by 
evaluating the diagram (3) of Fig.~\ref{fig:vertex} with the corrected propagator $D$, \begin{equation}
\tau^{-1}\equiv-2\text{Im}\Sigma\propto \omega^{d/3}.\label{oneloop2}
\end{equation}
Therefore, Landau's criterion $\omega\tau\rightarrow\infty$ as $\omega\rightarrow0$ does not hold when $d\leq3$, implying the breakdown of the FL theory.  

Thus, this one-loop treatment at least shows the instability of FLs and NGBs against infinitesimal couplings with $\vec{v}_{\vec{k},\vec{k}}\neq0$.  The ultimate fate of these interacting systems continues to be an active area of research~\cite{SSLee2,Sachdev1,Sachdev2,Senthil} and we do not expand further on that aspect here. We merely establish the condition when interactions with NGBs are relevant and render the decoupled fixed point unstable, similar to other well-studied cases.


Below, we demonstrate our general criterion through examples. 
\section{Example 1: Internal symmetries - conventional coupling}
Let us first discuss interactions between electrons and magnons in ferromagnets (in the absence of spin-orbit interactions).  
The coupling between the ferromagnetic order parameter $\vec{m}$ and the electron spin $\vec{s}=\psi^\dagger\frac{\vec{\sigma}}{2}\psi$ ($\vec{\sigma}$ is the Pauli matrix) may not contain any derivatives, e.g.,
\begin{eqnarray}
\mathcal{H}_{\text{el-magnon}}=J\vec{m}\cdot\vec{s}.
\end{eqnarray}
Hence it is not obvious that the electron-magnon vertex vanishes in the limit of small momentum transfer.  However, we know it must from our general criterion, as the spin $\vec{S}$ and the momentum $\vec{P}$ commute.  

To see this explicitly,  we perform a local $\text{SU}(2)$ rotation $U(\vec{x},t)$ defined by $U^\dagger(\vec{x},t)\hat{m}(\vec{x},t)\cdot\vec{\sigma}U(\vec{x},t)=\sigma_z$. Now, the spin-spin interaction becomes a site dependent Zeeman field along $s_z$, while electron-magnon interactions are included in derivatives of the rotated electron field $\partial_\mu \psi=U(\partial_\mu+i\mathcal{A}_\mu)\psi'$ through  $\mathcal{A}_\mu\equiv-i U^\dagger\partial_\mu U$.  If we expand $\mathcal{A}_\mu$ in series of NGB fields, each term contains one derivative acting on them.  Therefore, electron-magnon interactions vanish in the limit of small energy-momentum transfer (see~\cite{SM} for details).

In general, generators $Q_a$ of internal symmetries commute with $\vec{P}$ and therefore we always obtain {\em vanishing} couplings.

\section{Example 2: Continuous Space Rotation - nonvanishing coupling}
Our first nontrivial example is the spontaneous breaking of continuous spatial rotation symmetry. For concreteness, consider nematic order in $2+1$ dimensions, in which a circular Fermi surface is distorted into an ellipse in the ordered phase. The generator of $\text{SO}(2)$ spatial rotations is $L_z$, which does not commute with the momentum operator, $[L_z,P_i]=i\epsilon_{ij}P_j\neq0$ (where $i,\,j\in\{x,\,y\}$). Hence, from Eqn.~\ref{vertexkk}, we expect a nonvanishing coupling
\begin{eqnarray}
v_{\vec{k},\vec{k}}=\epsilon_{ij}\langle \vec{k}|p_j|\vec{k}\rangle \partial_{k_i}\epsilon_{\vec{k}}.\label{rotationv1}
\end{eqnarray} 
 where we assume a single band, and for simplicity, ignore spin. 
 
To see this from an explicit calculation, suppose that the spatial $\text{SO}(2)$ rotation is spontaneously broken by the order parameter $\langle\vec{n}\rangle=(1,0)^T$.   The Goldstone fluctuation $\theta$ of the order parameter $\vec{n}=(\cos\theta,\sin\theta)^T$ can couple to the spinless electron field $\psi$ via, e.g., $\mathcal{H}_{\text{int}}=(\chi/2m)|\vec{n}\cdot\vec{\nabla}\psi|^2$ as both $\nabla\psi$ and $\vec{n}$ are vector. Expanding the interaction to first order in $\theta$ ($\vec{n}\sim(1,\theta)^T$), we have
\begin{eqnarray}
\mathcal{H}_{\text{int}}^{(0)}&=&\chi\frac{\nabla_x\psi^\dagger\nabla_x\psi}{2m},\label{rotation0}\\
\mathcal{H}_{\text{int}}^{(1)}&=&\chi\theta\frac{\nabla_x\psi^\dagger\nabla_y\psi+\nabla_y\psi^\dagger\nabla_x\psi}{2m}.\label{rotation1}
\end{eqnarray}
Hence, the single-particle electron Hamiltonian is given by
\begin{eqnarray}
\mathcal{H}_0=\frac{\vec{p}^2}{2m}+\chi\frac{p_x^2}{2m}
\end{eqnarray}
leading to a single particle dispersion $\epsilon_{\vec{k}}=[(1+\chi)k_x^2+k_y^2]/2m$ for plane waves with wave vector $k$.  

By directly evaluating the matrix element of Eqn.~\ref{rotation1} for plane waves, we get 
\begin{equation}
v_{\vec{k}',\vec{k}}=\frac{\chi}{2m}(k_x' k_y+k_x k_y').\label{rotationv2}
\end{equation}
This is consistent with our criterion Eqn.~\ref{rotationv1}. Indeed, using the above dispersion $\epsilon_{\vec{k}}$ and $\langle n\vec{k}|p_j|n\vec{k}\rangle=k_j$ in Eqn.~\ref{rotationv1}, one gets $v_{\vec{k},\vec{k}}=\chi k_x k_y/m$, which agrees with Eqn.~\ref{rotationv2} in the limit $\vec{k}'\rightarrow\vec{k}$.

Note, the vertex does not vanish at $\vec{k}'=\vec{k}$ for generic points on the Fermi surface, except for few high-symmetry points with $k_x=0$ or $k_y=0$.  Therefore, at the most of the part of the Fermi surface, the quasi-particle lifetime is heavily suppressed by the interaction with the NGB $\theta$ originated from spontaneously broken continuous rotation.  A nematic order of a elliptically distorted Fermi surface~\cite{Oganesyan2001,Lawler2006} and a ferromagnetic order in the presence of a Rashba interaction~\cite{Cenke,Yasaman} are known examples of this mechanism.  

Finally, let us remark on a subtlety regarding space-time symmetries. In certain cases, even if the spatial rotation is spontaneously broken, NGBs associated with the broken rotation may not appear.  Suppose translations $p_{x,y}$ are spontaneously broken in $2+1$ dimensions.  Although rotation symmetry is also broken it does not lead to independent NGBs. Phonons originating from $p_{x,y}$ play the role of the NGB of $\ell_z$ as well, and the fluctuation $\theta$ associated with $\ell_z$ is related to displacement fields by $\theta=\partial_x u_y-\partial_y u_x$.  Although the field $\theta$ can couple strongly to electrons, these additional derivatives annihilate the scattering in the limit of small energy-momentum transfer.  Even when only $p_x$ or $p_y$ is broken, $\ell_z$ cannot produce an independent NGB. For example, helimagnets in $3+1$ dimensions with the spiral vector along the $z$-axis breaks $p_z-\ell_z$ and $\ell_{x,y}$ but the phonon associated with $p_z$ plays the role of NGBs of $\ell_{x,y}$ and orientational NGBs are absent~\cite{Radzihovsky,Low,WatanabeMurayama2,Hayata}. 


\section{Example 3: Magnetic Translation - nonvanishing coupling}
\begin{figure}
\begin{center}
\includegraphics[width=\columnwidth,clip]{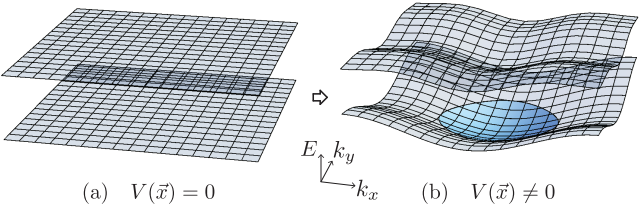}
\end{center}
\caption{(a) Electron band structure under a uniform magnetic field (Landau levels). (b) A spontaneously generated periodic lattice potential $V(\vec{x})$ produces dispersing bands.  The quasi-particle excitation of the partially-filled band (filled states shaded in blue) has a reduced life time due to the non vanishing electron-phonon interaction.}
\label{fig:band}
\end{figure}

As a new example of nonvanishing couplings, we discuss continuous translation under a uniform magnetic field in $2+1$ dimensions.  Suppose that a crystalline order with lattice vectors $\{\vec{a}_i\}_{i=1,2}$ is spontaneously formed, breaking the magnetic translation and giving birth to phonons (NGBs).   We assume an integer flux quantum per a unit cell for the commutativity of the lattice translations $T_i\equiv e^{i\vec{p}^B\cdot\vec{a}_i}$, where $(p_x^B,p_y^B)$ is given by $(-i\partial_x,\,-i\partial_y+eBx)$ in the Landau gauge $\vec{A}=-By\hat{x}$. Thanks to an effective periodic potential, the electron band structure becomes dispersive (Fig.~\ref{fig:band}). We are interested in coupling the NGBs (phonons) to quasiparticle excitations near the Fermi surface of a partially filled band.

In this case, the conserved (magnetic) momenta $\vec{p}^B$ also plays the role of broken generators $Q_a=p_a^B$ ($a=x,y$) that produce phonons.  Hence, we should look at the commutation relation $ [p_a^B,p_b^B]=-i\epsilon_{ab}eB\neq0$.  For discrete translations, we have $[\vec{p}^B,T_i]=-eB\hat{z}\times\vec{a}_iT_i$ (no sum over $i$), and we expect nonvanishing coupling from Eqns.~\ref{vanishing} and \ref{vertexdiscrete}:
\begin{equation}
\vec{v}_{n\vec{k},n\vec{k}}=eB\hat{z}\times\vec{a}_i\,(\vec{b}_i\cdot\vec{\nabla}_{\vec{k}}\epsilon_{n\vec{k}}).\label{magneticv1}
\end{equation}
Here $\{\vec{b}_i\}_{i=1,2}$ are reciprocal lattice vectors $\vec{a}_i\cdot\vec{b}_j=\delta_{ij}$.  Therefore, electrons may show NFL behaviors as a result of the nonvanishing interaction with phonons.  As we know, in the absence of the magnetic field $B=0$, the electron-phonon coupling is conventional as one can see from Eqn.~\ref{magneticv1}.

Let us confirm the nonvanishing coupling in Eqn.~\ref{magneticv1} from a direct calculation.   For simplicity, we assume one flux quantum per square lattice unit cell and assume the  mean-field lattice potential experienced by electrons is:
\begin{equation}
V(\vec{x})=-V_0\left[\cos(2\pi x/a)+\cos(2\pi y/a)\right].
\end{equation}

The electrons and phonons interact with each other through the potential $\mathcal{H}_{\text{int}}=V(\vec{x}-\vec{u})\psi^\dagger\psi$.  Expanding $\mathcal{H}_{\text{int}}$ in series of $\vec{u}$, we have
\begin{eqnarray}
\mathcal{H}_0&=&\frac{(\vec{p}-e\vec{A})^2}{2m}+V(\vec{x}),\\
\mathcal{H}_{\text{int}}^{(1)}&=&-\vec{u}\cdot\vec{\nabla}V.\label{magnetictranslation2}
\end{eqnarray}
We diagonalize $\mathcal{H}_0$ in the strong magnetic field limit, perturbatively taking into account the lattice potential to the lowest order in $mV_0/eB$.  In the Landau gauge, the lowest Landau level wave functions that simultaneously diagonalize $T_i$ are given by~\cite{Haldane,SM}
\begin{equation}
\Psi_{\vec{k}}\propto\sum_{m\in\mathbb{Z}}e^{-\frac{1}{2}\left(\frac{y}{\ell}+k_x\ell+\frac{2\pi \ell}{a}m\right)^2+i\left(k_x+\frac{2\pi}{a}m\right)x-ik_y am}.\label{landau}
\end{equation}
Therefore, the lowest electron band to first order in perturbation theory is
\begin{eqnarray}
\epsilon_{\vec{k}}&=&E_0+\langle\vec{k}|V|\vec{k}\rangle\notag\\
&=&\frac{eB}{2m}-\tilde{V}\left[\cos(k_ya)+\cos(k_xa)\right]\label{mph1}
\end{eqnarray}
with $\tilde{V}\equiv V_0\exp[-(\pi\ell/a)^2]$. Utilizing our criterion \ref{magneticv1}, and the dispersion above we predict the non vanishing coupling $\vec{v}_{\vec{k},\vec{k}}=eB\tilde{V}a\left (
-\sin {k_ya},\,
\sin {k_xa}\right )$ 

Now we directly computing the electron-phonon vertex by evaluating matrix elements of Eqn.~\ref{magnetictranslation2} with the zeroth-order wave function~\eqref{landau} and $u=u_{\vec{q}}e^{i\vec{q}\cdot\vec{x}}$, which gives:
\begin{equation}
\vec{v}_{\vec{k}',\vec{k}}=eB\tilde{V}a\,e^{-\frac{(q\ell)^2}{4}-iq_y\frac{k_x'+k_x}{2}\ell^2}
\begin{pmatrix}
-\sin \frac{(k_y'+k_y+iq_x)a}{2}\\
\sin \frac{(k_x'+k_x-iq_y)a}{2}
\end{pmatrix},\label{mph2}
\end{equation}
which in fact agrees with our formula when $\vec{k}'\rightarrow \vec{k}$.  


Let us now note some important physical consequences. For our results, it is important that  spontaneous breaking of translation symmetry occurs in a system with a uniform magnetic field. On the other hand, if the underlying symmetry is regular translation, and magnetic flux is spontaneously generated in the symmetry breaking process (as in a skyrmion lattice)  this does {\em not} lead to non-vanishing coupling~\cite{SkX}, and a Fermi liquid results. 
A different but equally valid viewpoint on our result is to consider the magnetic field being applied after breaking the translation symmetry (as in a crystal), which should modify the electron-phonon coupling. Therefore, in a clean metal, a magnetic field should induce a nonvanishing coupling between phonons and electrons. Although in principle this would have important consequences, in a typical solid, even at the highest available magnetic fields  there is a wide separation $\ell \gg a$ between the magnetic length $\ell=(eB)^{-{1/2}}$  and lattice spacing $a$. The typical dispersion of the Landau levels induced by the lattice, and hence the coupling constant is: $e^{-C(\ell/a)^2}\ll1$~\cite{SM}. Thus, although a non vanishing coupling is expected, its absolute magnitude is extremely small. A more promising physical scenario is the quantum Hall regime, where $\ell \sim a$, and where translation symmetry breaking into stripe and bubble phases are predicted and may have been observed in higher Landau levels~\cite{Koulakov,Fogler1,Moessner, Eisenstein}.  We leave an analysis of this interesting possibility to future work. 


\acknowledgements
We are grateful to Andrew Potter, Siddharth Parameswaran, Yasaman Bahri, Philipp Dumitrescu, and Tom\'a\v{s} Brauner for discussions and especially thank Max Metlitski for stimulating discussions at the early stages of this work and for useful comments on the draft.   H. W. was supported by the Honjo International Scholarship Foundation and A. V.  is supported by NSF - DMR 1206728.

\clearpage

\appendix
\onecolumngrid

\section{SUPPLEMENTAL MATERIAL\\
for ``Criterion for stability of Goldstone Modes and Fermi Liquid behavior in a metal with broken symmetry"}

\section{1. Interaction with constant NGB fields}
In this section, we prove the formula $\mathcal{H}_{\text{int}}^{(1)}=-[i\pi^aQ_a,\mathcal{H}_0]$ for constant NG fields $\pi^a$.  More precisely, we prove the Lagrangian version, $\mathcal{L}_{\text{int}}^{(1)}=-[i\pi^aQ_a,\mathcal{L}_0]$.  (Here, commutation relations with $Q_a$ and the Lagrangian density means the symmetry transformation of the fields contained in the Lagrangian, as we explain below.) These two statements are equivalent as long as symmetry generators $Q_i$ commute with the total Hamiltonian of the system. (We discuss few exceptions in the next section.)

In general, we can decompose the total Lagrangian density into three pieces,
\begin{eqnarray}
\mathcal{L}_{\text{tot}}=\mathcal{L}_{\text{el}}(\bar{\psi},\partial_\mu\bar{\psi},\psi,\partial_\mu\psi)+\mathcal{L}_{\text{NG}}(\pi^a,\partial_\mu\pi^a)+\mathcal{L}_{\text{int}}(\bar{\psi},\partial_\mu\bar{\psi},\psi,\partial_\mu\psi,\pi^a,\partial_\mu\pi^a).
\end{eqnarray}
We define $\mathcal{L}_{\text{int}}^{(1)}$ and $\mathcal{L}_0$ by
\begin{eqnarray}
\mathcal{L}_{\text{int}}^{(1)}&=&\left.\frac{\partial\mathcal{L}_{\text{int}}}{\partial\pi^a}\right|_{\pi=0}\pi^a+\left.\frac{\partial\mathcal{L}_{\text{int}}}{\partial\partial_\mu\pi^a}\right|_{\pi=0}\partial_\mu\pi^a,\\
\mathcal{L}_0&=&\mathcal{L}_{\text{el}}(\bar{\psi},\partial_\mu\bar{\psi},\psi,\partial_\mu\psi)+\mathcal{L}_{\text{int}}(\bar{\psi},\partial_\mu\bar{\psi},\psi,\partial_\mu\psi,0,0).
\end{eqnarray}
For constant $\pi^a$, we can drop the second term of $\mathcal{L}_{\text{int}}^{(1)}$ as $\partial_\mu\pi^a=0$.

\subsubsection{Internal symmetries}
Let us start with a general symmetry breaking pattern $G\rightarrow H$ of internal symmetries.  We introduce a NG field $\pi^a$ for each broken generator $Q_a$ ($a=1,2,\cdots,\mathrm{dim}G/H$) to describe low-energy fluctuations of the order parameter.  Under the symmetry transformation $U=e^{i\epsilon^iQ_i}$, NG fields transforms as 
\begin{eqnarray}
(\pi^a)'\equiv U_\epsilon \pi^aU_\epsilon^\dagger=\pi^a+\epsilon^ih_i^a(\pi)+O(\epsilon^2),
\end{eqnarray}
and its infinitesimal form is
\begin{eqnarray}
\delta_i \pi^a\equiv(\pi^a)'-\pi^a=[iQ_i,\pi^a]=h_i^a(\pi).
\end{eqnarray}
In the standard parametrization introduced by Refs.~[S. Coleman {\it et al}, Phys. Rev. 177, 2239 (1969), C. G. Callan {\it et al}, Phys. Rev. 177, 2247 (1969)], $h_b^a(\pi)=\delta_b^a+O(\pi)$ for broken generators $Q_b$ and $h_\rho^a(\pi)=O(\pi)$ for unbroken generators $Q_\rho$.  Namely, a broken generator $Q_a$ shifts $\pi^a$ by a constant and an unbroken generator $Q_\rho$ does not shift any NG fields by a constant amount.

Each term of the Lagrangian density $\mathcal{L}_{A}$ ($A=\text{el}, \text{NG}, \text{int}$) is invariant under the symmetry transformation (up to total derivatives).  Namely, $U_\epsilon \mathcal{L}_{A}U_\epsilon^\dagger=\mathcal{L}_{A}$.  Hence,
\begin{eqnarray}
0&=&{[iQ_a,\mathcal{L}_{\text{int}}]}\notag\\
&=&
[iQ_a,\bar{\psi}]\frac{\partial\mathcal{L}_{\text{int}}}{\partial\bar{\psi}}
+[iQ_a,\partial_\mu\bar{\psi}]\frac{\partial\mathcal{L}_{\text{int}}}{\partial\partial_\mu\bar{\psi}}
+[iQ_a,\psi]\frac{\partial\mathcal{L}_{\text{int}}}{\partial\psi}
+[iQ_a,\partial_\mu\psi]\frac{\partial\mathcal{L}_{\text{int}}}{\partial\partial_\mu\psi}
+[iQ_a,\pi^b]\frac{\partial\mathcal{L}_{\text{int}}}{\partial\pi^b}
+[iQ_a,\partial_\mu\pi^b]\frac{\partial\mathcal{L}_{\text{int}}}{\partial\partial_\mu\pi^b}.
\end{eqnarray}
We set $\pi^a=0$ {\it after}\/ substituting the relation $[iQ_b,\pi^a]=\delta_b^a+O(\pi)$:
\begin{eqnarray}
\left.\frac{\partial\mathcal{L}_{\text{int}}}{\partial\pi^a}\right|_{\pi=0}=-
\left(\left.[iQ_a,\bar{\psi}]\frac{\partial\mathcal{L}_{\text{int}}}{\partial\bar{\psi}}
+[iQ_a,\partial_\mu\bar{\psi}]\frac{\partial\mathcal{L}_{\text{int}}}{\partial\partial_\mu\bar{\psi}}
+[iQ_a,\psi]\frac{\partial\mathcal{L}_{\text{int}}}{\partial\psi}
+[iQ_a,\partial_\mu\psi]\frac{\partial\mathcal{L}_{\text{int}}}{\partial\partial_\mu\psi}\right)\right|_{\pi=0}
\end{eqnarray}
The right hand side is nothing but $-\left[iQ_a,\mathcal{L}_{\text{int}}|_{\pi=0}\right]$.  Hence, by multiplying $\pi^a$ to both hand side, we get
\begin{equation}
\left.\frac{\partial\mathcal{L}_{\text{int}}}{\partial\pi^a}\right|_{\pi=0}\pi^a=-\pi^a\left[iQ_a,\mathcal{L}_{\text{int}}|_{\pi=0}\right].
\end{equation}
Since $\mathcal{L}_{\text{el}}$ commutes with $Q_a$, we can add it to the inside of the commutator.  Therefore, for constant $\pi^a$, 
\begin{equation}
\mathcal{L}_{\text{int}}^{(1)}=-[i\pi^aQ_a,\mathcal{L}_0].
\end{equation}

As the simplest example, let us discuss the spin-spin interaction in ferromagnets.
\begin{eqnarray}
\mathcal{L}_{\text{el}}=i\psi^\dagger\partial_t\psi-\frac{|\vec{\nabla}\psi|^2}{2m},\quad \mathcal{L}_{\text{int}}=J\vec{n}\cdot\vec{s},
\end{eqnarray}
where $\vec{n}$ is the normalized ferromagnetic order parameter, $\psi$ is an electron field with the spin degree of freedom, $\vec{s}\equiv\psi^\dagger\vec{\sigma}\psi$ is the electron spin, and $\vec{\sigma}$ is the Pauli matrix.  The electron spin satisfies the commutation relation $[s_i(\vec{x},t),s_j(\vec{x}',t)]=i\epsilon_{ijk}s_k(\vec{x},t)\delta^d(\vec{x}-\vec{x}')$ and $[s_i(\vec{x},t),n_j(\vec{x}',t)]=0$.  

We introduce fluctuation $\pi_{x,y}(\vec{x},t)$ as $\vec{n}=(\pi_y,-\pi_x,1)^T+O(\pi_{x,y}^2)$.  By expanding the interaction to the linear order in fluctuation, we find
\begin{eqnarray}
\mathcal{L}_{\text{0}}&=&i\psi^\dagger\partial_t\psi-\frac{|\vec{\nabla}\psi|^2}{2m}+\frac{J}{2}s_z,\\
\mathcal{L}^{(1)}_{\text{int}}&=&J\left(\pi_ys_x-\pi_xs_y\right),
\end{eqnarray}
Then it can be readily shown that
\begin{equation}
\mathcal{L}_{\text{int}}^{(1)}=-\pi^x[iQ_x,\mathcal{L}_0]-\pi^y[iQ_y,\mathcal{L}_0]
\end{equation}
for $\vec{Q}=\int\mathrm{d}^dx\,(\vec{s}+m\vec{n})$.

Equivalently, in terms of the Hamiltonian,
\begin{eqnarray}
\mathcal{H}_0=\frac{\vec{p}^2}{2m}-Js_z,\quad \mathcal{H}_{\text{int}}^{(1)}=-J(\pi_ys_x-\pi_xs_y),
\end{eqnarray}
and it is straightforward to check
\begin{eqnarray}
\mathcal{H}_{\text{int}}^{(1)}=-\pi_x[iQ_x,\mathcal{H}_0]-\pi_y[iQ_y,\mathcal{H}_0].
\end{eqnarray}

\subsubsection{Translation}
Now we move on to spacetime symmetries.  As we will see, the above derivation applies with only some minor changes.  

Let us discuss translation $\vec{x}'=\vec{x}+\vec{a}$ as the easiest example.  The displacement field $\vec{u}(\vec{x},t)$ obeys the transformation rule,
\begin{eqnarray}
\vec{u}'(\vec{x},t)&\equiv& e^{i\vec{a}\cdot\vec{P}}\vec{u}(\vec{x},t)e^{-i\vec{a}\cdot\vec{P}}=\vec{u}(\vec{x}-\vec{a},t)+\vec{a},\\
\delta_j u^i(\vec{x},t)&\equiv&{u'}^i(\vec{x},t)-u^i(\vec{x},t)=[iP^j,u^i(\vec{x},t)]=\delta_j^i-\partial_ju^i.\label{SM:u}
\end{eqnarray}

Each component $\mathcal{L}_{A}$  is assumed to transforms as a scaler,
\begin{eqnarray}
\mathcal{L}_{A}'(\vec{x},t)&=&e^{i\vec{a}\cdot\vec{P}}\mathcal{L}_{A}(\vec{x},t)e^{-i\vec{a}\cdot\vec{P}}=\mathcal{L}_{A}(\vec{x}-\vec{a},t),\\
\delta\mathcal{L}_{A}(\vec{x},t)&=&[i\vec{P},\mathcal{L}_{A}(\vec{x},t)]=-\vec{\nabla}\mathcal{L}_{A}(\vec{x},t),
\end{eqnarray}
so that $L_{A}=\int\mathrm{d}^dx\,\mathcal{L}_{A}$ commutes with $\vec{P}$.  On the other hand, by explicitly computing $\delta\mathcal{L}_{\text{int}}=[i\vec{P},\mathcal{L}_{\text{int}}(\vec{x},t)]$, we have
\begin{eqnarray}
\delta\mathcal{L}_{\text{int}}&=&
[i\vec{P},\bar{\psi}]\frac{\partial\mathcal{L}_{\text{int}}}{\partial\bar{\psi}}
+[i\vec{P},\partial_\mu\bar{\psi}]\frac{\partial\mathcal{L}_{\text{int}}}{\partial\partial_\mu\bar{\psi}}
+[i\vec{P},\psi]\frac{\partial\mathcal{L}_{\text{int}}}{\partial\psi}
+[i\vec{P},\partial_\mu\psi]\frac{\partial\mathcal{L}_{\text{int}}}{\partial\partial_\mu\psi}
+[i\vec{P},u^i]\frac{\partial\mathcal{L}_{\text{int}}}{\partial u^i}
+[i\vec{P},\partial_\mu u^i]\frac{\partial\mathcal{L}_{\text{int}}}{\partial\partial_\mu u^i}.
\end{eqnarray}
Therefore, using Eq.~\eqref{SM:u} first, then setting $\vec{u}=0$ and multiplying $\vec{u}$, we get
\begin{eqnarray}
\left.\frac{\partial\mathcal{L}_{\text{int}}}{\partial \vec{u}}\right|_{\vec{u}=0}\cdot\vec{u}=-\vec{u}\cdot\left[i\vec{P},\mathcal{L}_{\text{int}}|_{\vec{u}=0}\right]-\vec{u}\cdot\vec{\nabla}\mathcal{L}_{\text{int}}|_{\vec{u}=0}.
\end{eqnarray}
Adding $0=-\vec{u}\cdot[i\vec{P},\mathcal{L}_{\text{el}}]-\vec{u}\cdot\vec{\nabla}\mathcal{L}_{\text{el}}$ to the right hand side, we get
\begin{equation}
\mathcal{L}_{\text{int}}^{(1)}=\left.\frac{\partial\mathcal{L}_{\text{int}}}{\partial \vec{u}}\right|_{\vec{u}=0}\cdot\vec{u}=-[i\pi^aQ_a,\mathcal{L}_0]-\vec{\nabla}\cdot(\vec{u}\mathcal{L}_0)
\end{equation}
for a constant NG field $\vec{u}$.  The last term is just a total derivative and can be dropped.

Even for the magnetic translation, this derivation does not change at all, since the displacement field is real and its transformation rule does not involve phase rotation. All characteristic features of the magnetic transformation are hidden in the commutation relation $[i\vec{P}_B,\psi]$.

\subsubsection{Rotation}
In the case of the spatial rotation $\vec{x}'=R_{\epsilon}\vec{x}$ with
\begin{eqnarray}
R_{\epsilon}&=&\begin{pmatrix}
\cos\epsilon&-\sin\epsilon\\
\sin\epsilon&\cos\epsilon,
\end{pmatrix},
\end{eqnarray}
the NG field originating from the rotational symmetry breaking transforms as
\begin{eqnarray}
\theta'(\vec{x},t)&\equiv& e^{i\epsilon L_z} \theta(\vec{x},t)e^{-i\epsilon L_z}=\theta(R_{-\epsilon}\vec{x},t)+\epsilon,\\
\delta \theta(\vec{x},t)&\equiv&\theta'(\vec{x},t)-\theta(\vec{x},t)=[iL_z,\theta(\vec{x},t)]=1-(x\partial_y-y\partial_x)\theta(\vec{x},t)
\end{eqnarray}
and the Lagranigian density transforms as
\begin{eqnarray}
\mathcal{L}_{A}'(\vec{x},t)&=&e^{i\epsilon L_z}\mathcal{L}_{A}(\vec{x},t)e^{-i\epsilon L_z}=\mathcal{L}_{A}(R_{-\epsilon}\vec{x},t),\\
\delta\mathcal{L}_{A}(\vec{x},t)&=&[iL_z,\mathcal{L}_{A}(\vec{x},t)]=-(x\partial_y-y\partial_x)\mathcal{L}_{A}(\vec{x},t)=-\partial_y(x\mathcal{L}_{A})+\partial_x(y\mathcal{L}_{A}),
\end{eqnarray}

In exactly the same way as above, we can show
\begin{eqnarray}
\left.\frac{\partial\mathcal{L}_{\text{int}}}{\partial\theta}\right|_{\theta=0}\theta=-\theta\left[iL_z,\mathcal{L}_0\right]-\theta\partial_y(x\mathcal{L}_0)+\theta\partial_x(y\mathcal{L}_0).
\end{eqnarray}
Again, when $\theta$ is constant, the last two terms are total derivatives and can be dropped.

\subsubsection{General spacetime symmetries}
It should be now clear how to extend the above derivation to general spacetime symmetries.    
NG fields $\pi^a$ corresponding to broken generators $Q_a$ obey
\begin{equation}
\delta\pi^a(\vec{x},t)=[iQ_a,\pi^b(\vec{x},t)]=\delta_a^b+f_a^\mu(\vec{x},t)\partial_\mu\pi^b(\vec{x},t)
\end{equation}
with some functions $f_a^\mu(\vec{x},t)$.
For those symmetries such that $\delta\mathcal{L}_{A}(\vec{x},t)=[iQ_a,\mathcal{L}_{A}(\vec{x},t)]$ is a total derivative, we have
\begin{eqnarray}
\left.\frac{\partial\mathcal{L}_{\text{int}}}{\partial\pi^a}\right|_{\pi=0}\pi^a=-\pi^a\left[iQ_a,\mathcal{L}_0\right]-\pi^a\partial_\mu F_a^\mu.
\end{eqnarray}
Hence, for constant $\pi^a$s, $\mathcal{L}_{\text{int}}^{(1)}=-[i\pi^aQ_a,\mathcal{L}_0]$ up to total derivatives.

\section{2. Photons as NGBs}
In this section, we discuss the electron-{\it photon}\/ interaction $\mathcal{L}_{\text{int}}=-A^\mu j_\mu$ in the framework developed above.  To that end, let us first review how to understand photons as NGBs.

The Lagrangian of the electron-photon interacting system is given by
\begin{eqnarray}
\mathcal{L}=i\bar{\psi}(\partial_t+ieA_t)\psi-\frac{|(\vec{\nabla}-ie\vec{A})\psi|^2}{2m}-\frac{1}{4}F_{\mu\nu}F^{\mu\nu}-\frac{1}{2\xi}(\partial_\mu A^\mu)^2.\label{SM:gauge}
\end{eqnarray}
Here, we added the term $-(1/2\xi)(\partial_\mu A^\mu)^2$ to fix the gauge (the $R_\xi$ gauge).  Then, the local gauge symmetry is reduced to the residual {\it global}\/ symmetry specified by
\begin{equation}
\psi=\psi e^{-ie\epsilon(x)},\quad A_\mu'=A_\mu+\partial_\mu\epsilon(x),\quad \epsilon(x)=a+b_\mu x^\mu,\quad a,b_\mu\in\mathbb{R}.
\end{equation}
Let $Q$ and $Q_\mu$ be the corresponding Noether charges and $P^\mu=(H,\vec{P})$ be the (four) momentum operator.  Their expressions are given by
\begin{eqnarray}
Q&=&\int\mathrm{d}^dx\,e\bar{\psi}\psi,\\
Q_\mu&=&\int\mathrm{d}^dx\left(\pi_\mu+ex_\mu\bar{\psi}\psi\right),\\
P_\mu&=&\int\mathrm{d}^dx\left[\pi_\nu\partial_\mu A^\nu+(i/2)(\bar{\psi}\partial_\mu\psi-\partial_\mu\bar{\psi}\psi)-\delta_{\mu0}\mathcal{L}\right],
\end{eqnarray}
where $[A^\mu(\vec{x},t),\pi_\nu(\vec{y},t)]=i\delta^\mu_\nu\delta^d(\vec{x}-\vec{y})$ and $[\psi(\vec{x},t),\bar{\psi}(\vec{y},t)]=\delta^d(\vec{x}-\vec{y})$.  Using these commutation relations, It is straightforward to verify that $[Q_\mu, A^\nu(\vec{x},t)]=-i\delta_\mu^\nu$ and $[Q_\mu,P^\nu]=-i\delta_\mu^\nu Q$

The relation $[Q_\mu, A^\nu(\vec{x},t)]=-i\delta_\mu^\nu$ indicates that $Q_\mu$'s are always broken.  As long as $Q$ is unbroken, they produce gapless bosons, which can be identified as photons.  $3-1=2$ transverse components are physical, while the longitudinal and the temporal component are unphysical.

When the $\text{U}(1)$ symmetry $Q$ is spontaneously broken, the NGB originating from $Q$ plays the role of those of $Q_\mu$. Hence, photons do not show up.  This is one way of understanding the Higgs phenomenon.  The $\text{U}(1)$ NGB does not appear in the physical spectrum either, since it belongs to the unphysical sector of the Hilbert space.

Let us now discuss the electron-photon scattering from this point of view.
The total Lagrangian density in Eq.~\eqref{SM:gauge} suggests that
\begin{eqnarray}
\mathcal{L}_{\text{el}}=0,\quad
\mathcal{L}_{\text{NG}}=-\frac{1}{4}F_{\mu\nu}F^{\mu\nu}-\frac{1}{2\xi}(\partial_\mu A^\mu)^2,\quad
\mathcal{L}_{\text{int}}=i\bar{\psi}(\partial_t+ieA_t)\psi-\frac{|(\vec{\nabla}-ie\vec{A})\psi|^2}{2m}.
\end{eqnarray}
Hence,
\begin{eqnarray}
\mathcal{L}_0=\mathcal{L}_{\text{int}}|_{A^\mu=0}=i\bar{\psi}\partial_t\psi-\frac{|\vec{\nabla}\psi|^2}{2m},\quad
\mathcal{L}_{\text{int}}^{(1)}=-A_0j^0+\vec{A}\cdot\vec{j}=-A_\mu j^\mu,
\end{eqnarray}
where $j^0=e\bar{\psi}\psi$ and $\vec{j}=(1/2mi)(\bar{\psi}\vec{\nabla}\psi-c.c.)$.  By the straightforward calculation, one can verify that 
\begin{equation}
-iA^\mu[Q_\mu,\mathcal{L}_0]=-A_\mu j^\mu=\mathcal{L}_{\text{int}}^{(1)}.
\end{equation}
Therefore, the electron-photon interaction can be understood as an example of the general class of electron-NGB interactions.

Note, however, that the Hamiltonian version of this relation $\mathcal{H}_{\text{int}}^{(1)}=A_\mu j^\mu=-iA^\mu[Q_\mu,\mathcal{H}_0]$ is not true for the temporal component $A_0$.  For the spatial component $\vec{A}$,  one can check it particularly easily in the single-particle picture, where $\hat{Q}_i=-e\hat{x}^i$, $\hat{H}_0=\hat{p}^2/2m$, and $-A^i[i\hat{Q}_i,\hat{H}_0]=ieA^i[\hat{x}^i,\hat{p}^2/2m]=A_i\hat{j}^i$ with $\hat{j}^i=e\hat{p}^i/m$.  However, $Q_0=e t$ ($t$ is not an operator) commutes with $\hat{H}_0$ and the linear interaction $eA_0$ cannot be written as $-A_0[i\hat{Q}_0,\hat{H}_0]$.  This failure originates from the fact that $Q_0$ {\it does not}\/ commutes with Hamiltonian $[Q_0,H]=-iQ\neq0$, even though $Q_0$ is still a symmetry of the system in the sense that it leaves the Lagrangian invariant.  

Another example of this type is the Galilean boost $\vec{B}$, which satisfies $[\vec{B}, H]=-i\vec{P}\neq0$.  So if the Galilean boost was spontaneously broken and if it produced independent NGBs $\vec{v}$, their coupling to electrons would not satisfy the relation $\mathcal{H}_{\text{int}}^{(1)}=-i\vec{v}\cdot[\vec{B},\mathcal{H}_0]$.  However, the Galilean boost does not usually produce NGBs. For example, in superfluids, the Bogoliubov mode originating from spontaneously broken $\text{U}(1)$ symmetry plays the role of the NGB of the boost symmetry, again due to the linear dependence of the current operator.

\section{3. Singularities in the matrix element $\langle n\vec{k}'|Q_a|n\vec{k}\rangle$}
When an operator $Q_a$ does not commute with the generator of the translation $\vec{P}$, {\it i.e.}\/, $[Q_a,\vec{P}]\neq0$, we have $[Q_a,e^{i\vec{P}\cdot\vec{a}_i}]\neq0$.  By further assuming that $\langle n\vec{k}|[Q_a,e^{i\vec{P}\cdot\vec{a}_i}]|n\vec{k}\rangle\neq0$, which is generically true except for some high symmetry points in the Brillouin zone, one can prove that the expectation value $\langle n\vec{k}|Q_a|n\vec{k}\rangle$ is not well-defined.

For example, using commutation relations
\begin{eqnarray}
{[x_i,p_j]}&=&i\delta_{ij},\\
{[\ell_z,p_i]}&=&i\epsilon_{ij}p_j,\\
{[p_i^B,p_j^B]}&=&-i\epsilon_{ij}eB,
\end{eqnarray}
one can show 
\begin{eqnarray}
{[\vec{x},e^{i\vec{p}\cdot\vec{a}_i}]}&=&-\vec{a}_i\,e^{i\vec{p}\cdot\vec{a}_i},\\
{[\ell_z,e^{i\vec{p}\cdot\vec{a}_i}]}&=&-\hat{z}\cdot\vec{a}_i\times\vec{p}\,e^{i\vec{p}\cdot\vec{a}_i},\\
{[\vec{p}^B,e^{i\vec{p}^B\cdot\vec{a}_i}]}&=&-eB\hat{z}\times\vec{a}_i\,e^{i\vec{p}^B\cdot\vec{a}_i},
\end{eqnarray}
respectively.  We now evaluate the matrix element of these commutation relations using the definition $e^{i\vec{p}\cdot\vec{a}_i}|n\vec{k}\rangle=e^{i\vec{k}\cdot\vec{a}_i}|n\vec{k}\rangle$. One then finds
\begin{eqnarray}
\langle n'\vec{k}'|\vec{x}|n\vec{k}\rangle&=&-\frac{e^{i\vec{k}\cdot\vec{a}_i}}{e^{i\vec{k}\cdot\vec{a}_i}-e^{i\vec{k}'\cdot\vec{a}_i}}\vec{a}_i\delta_{\vec{k}',\vec{k}}\delta_{n',n}\simeq\frac{i\vec{a}_i}{(\vec{k}-\vec{k'})\cdot\vec{a}_i}\delta_{\vec{k}',\vec{k}}\delta_{n',n}+O((\vec{k}-\vec{k}')^0),\\
\langle n'\vec{k}'|\ell_z|n\vec{k}\rangle&=&-\frac{e^{i\vec{k}\cdot\vec{a}_i}}{e^{i\vec{k}\cdot\vec{a}_i}-e^{i\vec{k}'\cdot\vec{a}_i}}\hat{z}\cdot\vec{a}_i\times\vec{k}\delta_{\vec{k}',\vec{k}}\delta_{n',n}=\frac{i\hat{z}\cdot\vec{a}_i\times\vec{k}}{(\vec{k}-\vec{k'})\cdot\vec{a}_i}\delta_{\vec{k}',\vec{k}}\delta_{n',n}+O((\vec{k}-\vec{k}')^0),\\
\langle n'\vec{k}'|\vec{p}^B|n\vec{k}\rangle&=&-\frac{e^{i\vec{k}\cdot\vec{a}_i}}{e^{i\vec{k}\cdot\vec{a}_i}-e^{i\vec{k}'\cdot\vec{a}_i}}eB\hat{z}\times\vec{a}_i\delta_{\vec{k}',\vec{k}}\delta_{n',n}=\frac{ieB\hat{z}\times\vec{a}_i}{(\vec{k}-\vec{k'})\cdot\vec{a}_i}\delta_{\vec{k}',\vec{k}}\delta_{n',n}+O((\vec{k}-\vec{k}')^0).
\end{eqnarray}
This is how one usually derives $\langle x|\hat{p}|x'\rangle=-i\hbar\delta(x-x')\partial_{x'}$ in the single-particle quantum mechanics.

\section{4. Comoving frame of NGBs}
In the main text, we discuss the property of electron-NGB vertices using commutation relations.  In this section we discuss them from an alternative approach.
\subsection{a. Magnons in ferromagnets}
The spin-spin interaction in ferromagnetic metals reads
\begin{equation}
\mathcal{H}_{\text{int}}=-\frac{J}{2}\vec{n}\cdot \bar{\psi}\vec{\sigma}\psi.
\end{equation}
It is not obvious from this representation that the electron-magnon vertex vanishes in the limit of small momentum transfer, since the NGB fields in this interaction does not contain derivatives acting on them.  However, there is a useful trick to convert these non-derivative interactions into those with at least one derivative acting on NGB fields. Namely, we perform a local $\text{SU}(2)$ rotation $U(\vec{x},t)$ defined by $U^\dagger(\vec{x},t)\vec{n}(\vec{x},t)\cdot\vec{\sigma}U(\vec{x},t)=\sigma_z$. In other words, we take the quantization axis of the electron spin in the comoving frame of the ferromagnetic order parameter.  The spin-spin interaction in terms of the new field $\psi'=U^{-1}\psi$ becomes a constant spin-dependent chemical potential $J{\psi'}^\dagger\sigma_z\psi'/2$.  Electron-magnon interactions are instead included in derivatives of the electron field $\partial_\mu \psi=U(\partial_\mu+i\mathcal{A}_\mu)\psi'$ through fluctuations of the Berry phase $\mathcal{A}_\mu\equiv-i U^\dagger\partial_\mu U$.  If we expand $\mathcal{A}_\mu$ in series of NGB fields, each term contains one derivative acting on them.  Therefore, electron-magnon interactions in $i{\psi'}^\dagger(\partial_t-i\mathcal{A}_0)\psi'$ and $[(\vec{\nabla}-i\vec{\mathcal{A}})\psi']^2$ vanish in the limit of small energy-momentum transfer.

\subsection{b. Phonons in crystals}
Similarly, the electron-phonon interaction in
\begin{eqnarray}
\mathcal{H}_{\text{int}}=V(\vec{x}-\vec{u})\bar{\psi}(\vec{x},t)\psi(\vec{x},t)
\end{eqnarray}
does not contain derivatives acting on the displacement field $\vec{u}(\vec{x},t)$, but the electron-phonon scattering vanishes in the limit of small energy-momentum transfer as discussed by using commutation relations.  

To see the vanishing scattering more clearly, we can convert the non-derivative coupling $V(\vec{x}-\vec{u})$ into derivative ones by going to the comoving frame of the crystal lattice.  That is, we change the integration variable of the Lagrangian from $\vec{x}$ to $\vec{x}'=\vec{x}-\vec{u}$ and redefine the electron field $\psi'(\vec{x}',t)=\psi(\vec{x},t)$.  Then the potential $V(\vec{x}-\vec{u})=V(\vec{x}')$ can no longer fluctuate, analogously to the above spin-spin interaction after the SU(2) rotation.  Instead, all the electron-phonon interactions come from rewriting the volume element and derivatives:
\begin{eqnarray}
\mathrm{d}^dx\mathrm{d}t&=&\mathrm{d}^dx'\mathrm{d}t'(1+\vec{\nabla}'\cdot\vec{u})+O((\partial \vec{u})^2),\\
\partial_\mu&=&\partial_\mu'-(\partial_\mu'u^i)\partial_i'+O((\partial \vec{u})^2).
\end{eqnarray}
It is now clear in this representation that all electron-phonon interactions vanish for a constant $\vec{u}$.

\subsection{c. Orientational NGBs in phases with rotational symmetry breaking}
If we can eliminate all non-derivative couplings by going to the comoving frame of NGBs, there is no hope to get non-vanishing couplings, as derivatives on NGBs vanish in the limit of small energy-momentum transfer.  Here we discuss why this comoving frame argument fails in the case of sptial rotation and magnetic translation. (More generally, spacetime symmetries except for the ordinary translation.)

If possible, we would like to eliminate all non-derivative couplings in the interacting Lagrangian,
\begin{equation}
\int\mathrm{d}^dx\mathrm{d}t|\vec{n}\cdot\vec{\nabla}\psi|^2=\int\mathrm{d}^dx\mathrm{d}t
\begin{pmatrix}
\cos\theta\\
\sin\theta
\end{pmatrix}
\cdot\vec{\nabla}\bar{\psi}
\begin{pmatrix}
\cos\theta\\
\sin\theta
\end{pmatrix}
\cdot\vec{\nabla}\psi.
\end{equation}
If we change the integration variable from $\vec{x}$ to $\vec{x}'=R_{\epsilon}\vec{x}$, we get
\begin{equation}
\int\mathrm{d}^dx'\mathrm{d}t
\begin{pmatrix}
\cos(\theta-\epsilon)\\
\sin(\theta-\epsilon)
\end{pmatrix}
\cdot\vec{\nabla}'\bar{\psi}
\begin{pmatrix}
\cos(\theta-\epsilon)\\
\sin(\theta-\epsilon)
\end{pmatrix}
\cdot\vec{\nabla}'\psi,
\end{equation}
where 
\begin{eqnarray}
R_{\epsilon}&=&\begin{pmatrix}
\cos\epsilon&-\sin\epsilon\\
\sin\epsilon&\cos\epsilon
\end{pmatrix}
\end{eqnarray}
is the orthogonal matrix for the rotation by a {\it constant}\/ angle $\epsilon$.  Therefore, changing the integration variable effectively shifts $\theta$ by $-\epsilon$. Thus one may expect that setting $\epsilon(\vec{x},t)=\theta(\vec{x},t)$ locally eliminates all $\theta$ dependence without derivatives.  However, it does not work for the following reason.  If we define $\vec{x}'=R_{\theta(\vec{x},t)}\vec{x}$ and rewrite derivative $\vec{\nabla}$ in terms of $\vec{\nabla}'$, we find
\begin{equation}
\partial_i=(\partial_ix'^j)\partial_j'=\partial_i[(R_{\theta})^j_{\,\,k}x^k]\partial_j'=(R_{\theta})^j_{\,\,i}\partial_j'+(\partial_iR_{\theta})^j_{\,\,k}x^k\partial_j'.
\end{equation}
Due to the second term of the last expression, the Lagrangian now explicitly depends on the coordinate.  This makes the Lagrangian after the rotation completely useless for any realistic calculations. Especially, we cannot use the Fourier transformation (despite the fact that the translation is not actually broken).  Therefore, we cannot discuss the behavior of couplings in the limit of the small momentum transfer.

\subsection{d. Magnetic translation}
We now discuss the magnetic translation.  We would like to remove $\vec{u}$ without derivatives in the Lagrangian,
\begin{eqnarray}
\mathcal{L}_{\text{el$+$int}}=i\bar{\psi}\partial_t\psi-\frac{|(\vec{\nabla}-ie\vec{A})\psi|^2}{2m}-\bar{\psi}\psi V(\vec{x}-\vec{u}).
\end{eqnarray}
If we just change the integration variable to $\vec{x}'=\vec{x}-\vec{u}(\vec{x},t)$, then $\vec{u}$ without derivatives appears from the vector potential,
\begin{equation}
\vec{A}=B\begin{pmatrix}-y\\0\\0\end{pmatrix}=B\begin{pmatrix}-y'-u_y\\0\\0\end{pmatrix}.
\end{equation}
In order to absorb this new $\vec{u}$ dependence, one can further perform a local gauge transformation,
\begin{equation}
\psi'=e^{-ieB x' u_y}\psi.
\end{equation}
When $u_y$ is a constant, this combination of the translation and the gauge transformation successfully removes all $u_y$'s from the Lagrangian. However, for a general $u_y(\vec{x},t)$, we have
\begin{equation}
\vec{\nabla}'\psi'=e^{-ieB x' u_y}\left(\vec{\nabla}'\psi-ieB\hat{x}u_y\psi-ieBx'\psi\vec{\nabla}'u_y\right).
\end{equation}
Again the last term introduces an undesirable coordinate dependence to the Lagrangian.

\section{5. Landau levels with lattice momentum}
In this section, we summarize the wave function of Landau levels (following Ref.~[F. D. M. Haldane and E. H. Rezayi, Phys. Rev. B 31, 2529 (1985)]) that simultaneously diagonalize Hamiltonian and lattice translations,
\begin{eqnarray}
H=\frac{(p_x+eB y)^2+p_y^2}{2m},\quad T_x=e^{ip_xa_x},\quad T_y=e^{i(p_y+eB x)a_y}.
\end{eqnarray}
We assume a rectangular lattice with primitive lattice vectors $\vec{a}_x=a_x\hat{x}$ and $\vec{a}_y=a_y\hat{y}$ and a flux quantum per a unit cell $eBa_xa_y=2\pi$.  We work in a torus $a_xN_x\times a_y N_y$ ($N_x, N_y\in\mathbb{Z}$) and impose the periodic boundary condition $T_x^{N_x}=T_y^{N_y}=1$.  The number of degeneracy is precisely the number of lattice points,
\begin{equation}
\frac{a_xa_yN_xN_y}{2\pi \ell^2}=N_xN_y.\quad \ell\equiv\sqrt{\frac{1}{eB}}.
\end{equation}
For each $k=\frac{2\pi}{a_xN_x}i\quad(i=1,2,\cdots,N_xN_y)$, the function
\begin{eqnarray}
\psi_{nk}(\vec{x})
=\sum_{j\in\mathbb{Z}}
\frac{H_n\left(\frac{y}{\ell}+k\ell+\frac{2\pi \ell}{a_x}jN_y\right)
e^{-\frac{1}{2}\left(\frac{y}{\ell}+k\ell+\frac{2\pi \ell}{a_x}jN_y\right)^2}}{\sqrt{2^nn!\sqrt{\pi}\ell}}\frac{e^{i\left(k+\frac{2\pi}{a_x}jN_y\right)x}}{\sqrt{a_xN_x}}
\end{eqnarray}
represents an simultaneous eigenfunction of the Hamiltonian with the eigenvalue $(eB/m)(n+1/2)$ and the lattice translation $T_x$:
\begin{eqnarray}
T_x\psi_{nk}(\vec{x})
=\sum_{j\in\mathbb{Z}}
\frac{H_n\left(\frac{y}{\ell}+k\ell+\frac{2\pi \ell}{a_x}jN_y\right)
e^{-\frac{1}{2}\left(\frac{y}{\ell}+k\ell+\frac{2\pi \ell}{a_x}jN_y\right)^2}}{\sqrt{2^nn!\sqrt{\pi}\ell}}\frac{e^{i\left(k+\frac{2\pi}{a_x}jN_y\right)(x+a_x)}}{\sqrt{a_xN_x}}=e^{ika_x}\psi_{nk}(\vec{x}).
\end{eqnarray}
In order to make it a simultaneous eigenfunction of $T_y$ as well, we take a superposition
\begin{eqnarray}
\Psi_{n\vec{k}}(\vec{x})&\equiv&\sum_{m=1}^{N_y}\frac{e^{-ik_y a_ym}}{\sqrt{N_y}}\psi_{n,k_x+\frac{2\pi}{a_x}m}(\vec{x})\notag\\
&=&\sum_{m=1}^{N_y}\sum_{j\in\mathbb{Z}}\frac{e^{-ik_y a_y(m+jN_y)}}{\sqrt{N_y}}
\frac{H_n\left(\frac{y}{\ell}+k_x\ell+\frac{2\pi \ell}{a_x}(m+jN_y)\right)
e^{-\frac{1}{2}\left(\frac{y}{\ell}+k_x\ell+\frac{2\pi \ell}{a_x}(m+jN_y)\right)^2}}{\sqrt{2^nn!\sqrt{\pi}\ell}}\frac{e^{i\left(k_x+\frac{2\pi}{a_x}(m+jN_y)\right)x}}{\sqrt{a_xN_x}}\notag\\
&=&\sum_{m\in\mathbb{Z}}\frac{e^{-ik_y a_ym}}{\sqrt{N_y}}
\frac{H_n\left(\frac{y}{\ell}+k_x\ell+\frac{2\pi \ell}{a_x}m\right)
e^{-\frac{1}{2}\left(\frac{y}{\ell}+k_x\ell+\frac{2\pi \ell}{a_x}m\right)^2}}{\sqrt{2^nn!\sqrt{\pi}\ell}}\frac{e^{i\left(k_x+\frac{2\pi}{a_x}m\right)x}}{\sqrt{a_xN_x}},
\end{eqnarray}
where 
\begin{eqnarray}
k_x&=&\frac{2\pi}{a_x}i_x \quad i_x=1,2,\cdots,N_x,\\
k_y&=&\frac{2\pi}{a_y}i_y \quad i_y=1,2,\cdots,N_y.
\end{eqnarray}

Now $\Psi_{n\vec{k}}(\vec{x})$'s are simultaneous eigenstates of $T_y$ as well:
\begin{eqnarray}
T_y\Psi_{n\vec{k}}(\vec{x})
&=&e^{ieB xa_y}\sum_{m\in\mathbb{Z}}\frac{e^{-ik_y a_ym}}{\sqrt{N_y}}
\frac{H_n\left(\frac{y+a_y}{\ell}+k_x\ell+\frac{2\pi \ell}{a_x}m\right)
e^{-\frac{1}{2}\left(\frac{y+a_y}{\ell}+k_x\ell+\frac{2\pi \ell}{a_x}m\right)^2}}{\sqrt{2^nn!\sqrt{\pi}\ell}}\frac{e^{i\left(k_x+\frac{2\pi}{a_x}m\right)x}}{\sqrt{a_xN_x}}\notag\\
&=&\sum_{m\in\mathbb{Z}}\frac{e^{-ik_y a_ym}}{\sqrt{N_y}}
\frac{H_n\left(\frac{y}{\ell}+k_x\ell+\frac{2\pi \ell}{a_x}(m+1)\right)
e^{-\frac{1}{2}\left(\frac{y}{\ell}+k_x\ell+\frac{2\pi \ell}{a_x}(m+1)\right)^2}}{\sqrt{2^nn!\sqrt{\pi}\ell}}\frac{e^{i\left(k_x+\frac{2\pi}{a_x}(m+1)\right)x}}{\sqrt{a_xN_x}}\notag\\
&=&e^{ik_ya_y}\Psi_{n\vec{k}}(\vec{x}).
\end{eqnarray}
For the lowest Landau levels, we have
\begin{eqnarray}
\Psi_{\vec{k}}(\vec{x})\equiv \Psi_{0\vec{k}}(\vec{x})=\sum_{m\in\mathbb{Z}}
\frac{e^{-\frac{1}{2}\left(\frac{y}{\ell}+k_x\ell+\frac{2\pi \ell}{a_x}m\right)^2+i\left(k_x+\frac{2\pi}{a_x}m\right)x-ik_y a_ym}}{\sqrt{\sqrt{\pi}\ell a_xN_xN_y}}.
\end{eqnarray}

\section{6. The electron Green function under magnetic field}
Here we summarize the free electron Green function under the magnetic field.  We expand the electron field operator as
\begin{equation}
\psi(\vec{x},t)=\sum_{n\vec{k}}\psi_{n\vec{k}}(\vec{x})c_{n\vec{k}}(t),
\end{equation}
where $c_{n\vec{k}}(t)$ is the annihilation operator of electrons in the Bloch eigenstate $\psi_{n\vec{k}}(\vec{x})$, either with or without an external magnetic field.  $c_{n\vec{k}}(t)$'s satisfy the equal-time anti-commutation relation
\begin{equation}
\{c_{n\vec{k}}(t),c_{n'\vec{k}'}^\dagger(t)\}=\delta_{n n'}\delta_{\vec{k},\vec{k}'}.  
\end{equation}
The free Hamiltonian can be expressed as
\begin{equation}
H_0=\sum_{n\vec{k}}\epsilon_{n\vec{k}}c_{n\vec{k}}^\dagger c_{n\vec{k}}.
\end{equation}
Thus the time-evolution of the annihilation operator under $H_0$ is $c_{n\vec{k}}(t)=c_{n\vec{k}}e^{-i\epsilon_{n\vec{k}}t}$.  The free Green function is then given by
\begin{eqnarray}
G_n(\vec{k},t)&\equiv&-i\langle T c_{n\vec{k}}(t)c_{n\vec{k}}^\dagger(0)\rangle\notag\\
&=&-i\langle c_{n\vec{k}}c_{n\vec{k}}^\dagger\rangle e^{-i\epsilon_{n\vec{k}}t}\theta(t)+i\langle c_{n\vec{k}}^\dagger c_{n\vec{k}}\rangle e^{-i\epsilon_{n\vec{k}}t}\theta(-t)\notag\\
&=&\int\frac{\mathrm{d}\omega}{2\pi}e^{-i\omega t} \left[\frac{\theta(\epsilon_{n\vec{k}})}{\omega-\epsilon_{n\vec{k}}+i\delta}+\frac{\theta(-\epsilon_{n\vec{k}})}{\omega-\epsilon_{n\vec{k}}-i\delta}\right]\notag\\
&\equiv&\int\frac{\mathrm{d}\omega}{2\pi}e^{-i\omega t} G_{n}(\vec{k},\omega).
\end{eqnarray}
In the derivation, we assumed that single-particle states with $\epsilon_{n\vec{k}}<0$ are filled and otherwise unfilled.  Therefore, the electron Green function in the momentum space takes the same form regardless of the presence or absence of the external magnetic field.

\section{7. Cancelation of the induced mass of NGBs}
For completeness, here we check the absence of a mass of NGBs generated by integrating out electrons.  
\begin{figure}[b!]
\begin{center}
\includegraphics[width=0.15\columnwidth,clip]{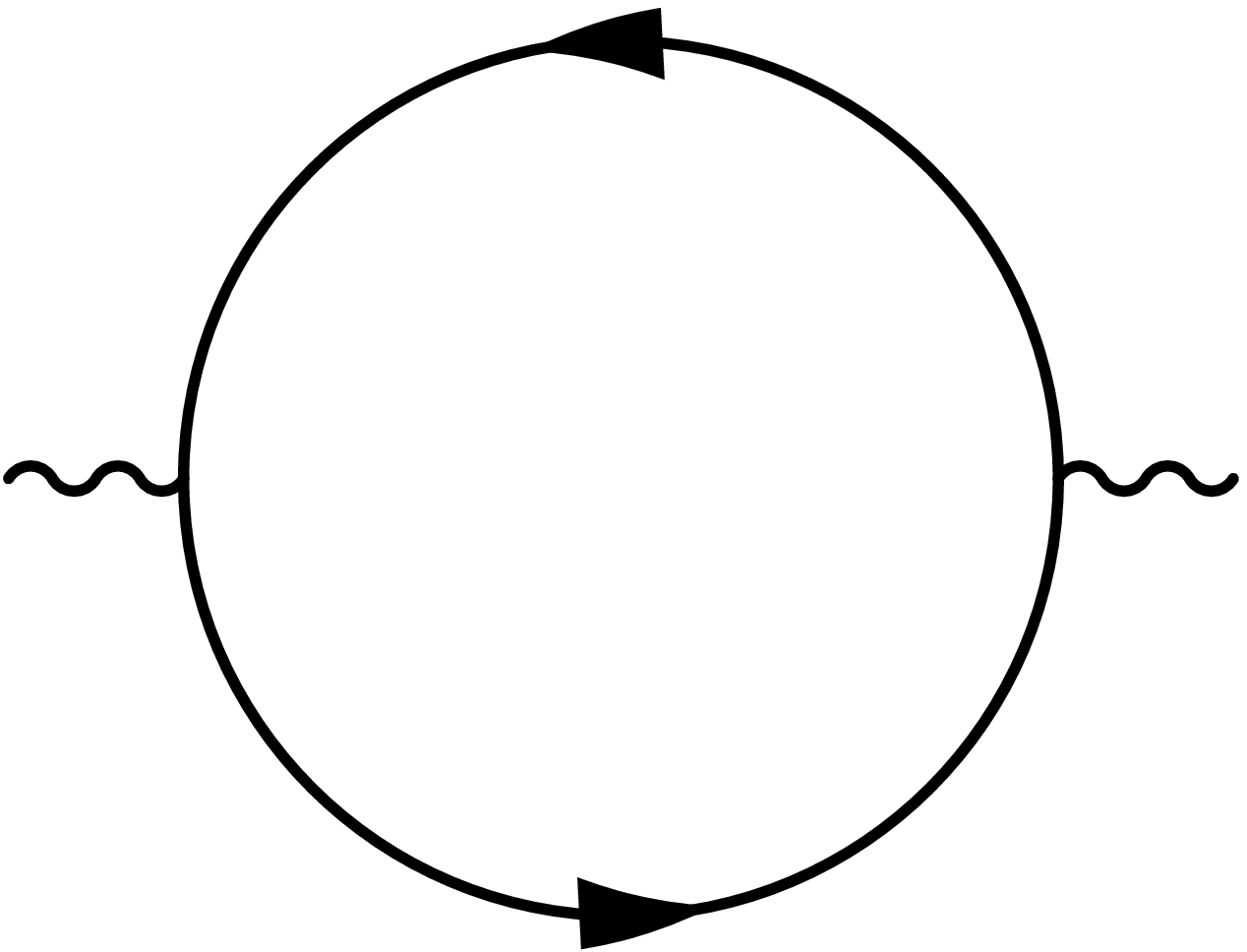}\quad\quad\quad
\includegraphics[width=0.08\columnwidth,clip]{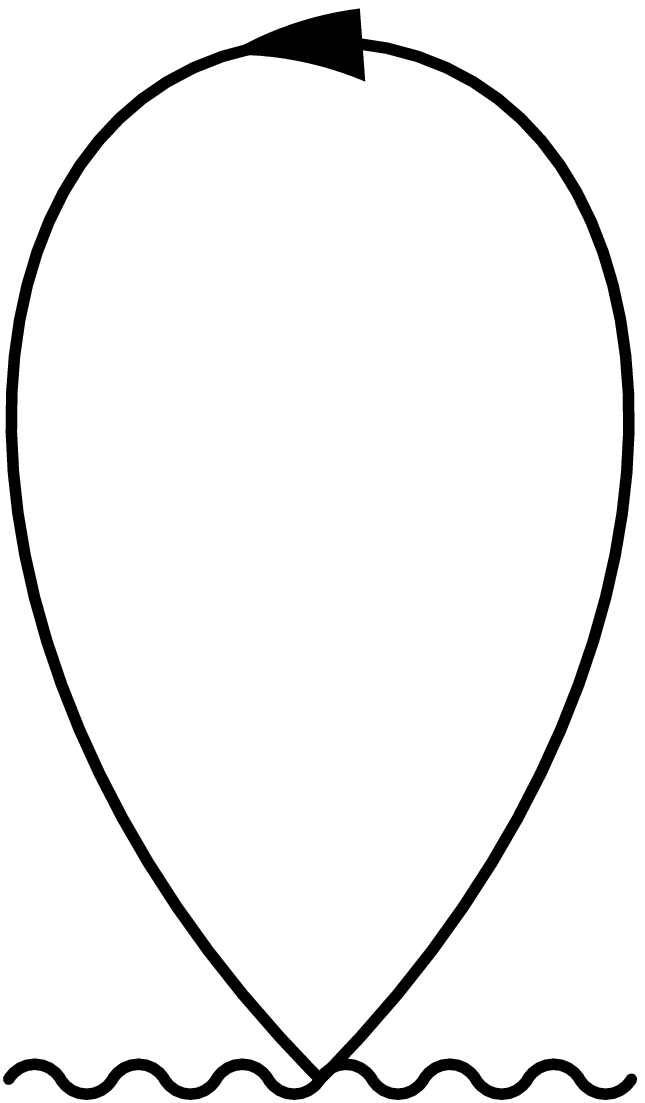}
\end{center}
\caption{1-loop diagrams for boson self-energy corrections.}
\label{fig:boson}
\end{figure}
\subsection{a. Rotation}
Let us start with the example of the spatial rotation.  For a constant $\theta$, we have
\begin{eqnarray}
H_{\text{int}}&=&\frac{\chi}{2m}\left[(k_x\cos\theta+k_y\sin\theta)^2-k_x^2\right]\psi_k^\dagger\psi_k\notag\\
&=&\frac{\chi}{m}\left[\theta k_xk_y+\frac{1}{2}\theta^2(k_y^2-k_x^2)+O(\theta^3)\right]\psi_k^\dagger\psi_k\notag\\
&=&\left[-\theta\partial_{\phi_{\vec{k}}}\epsilon_{\vec{k}}+\frac{1}{2}\theta^2\partial_{\phi_{\vec{k}}}^2\epsilon_{\vec{k}}+O(\theta^3)\right]\psi_k^\dagger\psi_k,
\end{eqnarray}
where $k e^{i\phi_{\vec{k}}}=k_x+ik_y$ and $\epsilon_{\vec{k}}$ is the electron dispersion,
\begin{equation}
\epsilon_{\vec{k}}=\frac{(1+\chi)k_x^2+k_y^2}{2m}-\mu. 
\end{equation}
The boson self-energy $\Pi$ at $\vec{q}=0$ and $\nu=0$ has two contributions at the 1-loop level,
\begin{equation}
\Pi(0)=\int\frac{\mathrm{d}^2k\mathrm{d}\omega}{(2\pi)^3}\left[\left(\partial_{\phi_{\vec{k}}}\epsilon_{\vec{k}}\,G(\vec{k},\omega)\right)^2+\partial_{\phi_{\vec{k}}}^2\epsilon_{\vec{k}}\,G(\vec{k},\omega)\right].
\end{equation}
Here the first (second) term represents the left (right) diagram in Fig.~\ref{fig:boson}. To show their cancelation, we use the relation of the electron Green function $G^{-1}(\vec{k},\omega)=\omega-\epsilon_{\vec{k}}$:
\begin{equation}
\vec{\nabla}_{\vec{k}}G(\vec{k},\omega)=[G(\vec{k},\omega)]^2\vec{\nabla}_{\vec{k}}\epsilon_{\vec{k}}.\label{app:delG}
\end{equation}
Then,
\begin{eqnarray}
\Pi(0)&=&\int\frac{\mathrm{d}^2k\mathrm{d}\omega}{(2\pi)^3}\left[\partial_{\phi_{\vec{k}}}\epsilon_{\vec{k}}\,\partial_{\phi_{\vec{k}}}G(\vec{k},\omega)+\partial_{\phi_{\vec{k}}}^2\epsilon_{\vec{k}}\,G(\vec{k},\omega)\right]\notag\\
&=&\int\frac{\mathrm{d}^2k\mathrm{d}\omega}{(2\pi)^3}\left[-\partial_{\phi_{\vec{k}}}^2\epsilon_{\vec{k}}\,G(\vec{k},\omega)+\partial_{\phi_{\vec{k}}}^2\epsilon_{\vec{k}}\,G(\vec{k},\omega)\right]=0.
\end{eqnarray}

\subsection{b. Magnetic translation}
Next, for the electron-phonon problem under a magnetic field, we have
\begin{eqnarray}
H_{\text{int}}&=&-\tilde{V}_x\left[\cos\big(k_ya_y-\frac{2\pi}{a_x}u_x\big)-\cos\big(k_ya_y\big)\right]-\tilde{V}_y\left[\cos\big(k_xa_x+\frac{2\pi}{a_y}u_y\big)-\cos\big(k_xa_x\big)\right]\notag\\
&=&\left[\left(\frac{2\pi u_y}{a_y}\right)\partial_{k_xa_x}\epsilon_{\vec{k}}-\left(\frac{2\pi u_x}{a_x}\right)\partial_{k_ya_y}\epsilon_{\vec{k}}\right]+\frac{1}{2}\left[\left(\frac{2\pi u_y}{a_y}\right)^2\partial_{k_xa_x}^2\epsilon_{\vec{k}}+\left(\frac{2\pi u_x}{a_x}\right)^2\partial_{k_ya_y}^2\epsilon_{\vec{k}}\right]+O(u^3).
\end{eqnarray}
Therefore, again by using Eq.~\eqref{app:delG},
\begin{eqnarray}
\Pi_{xx}(0)&=&\left(\frac{2\pi }{a_x}\right)^2\int\frac{\mathrm{d}^2k\mathrm{d}\omega}{(2\pi)^3}\left[\left((\partial_{k_ya_y}\epsilon_{\vec{k}})G(\vec{k},\omega)\right)^2+(\partial_{k_ya_y}^2\epsilon_{\vec{k}})G(\vec{k},\omega)\right]\notag\\
&=&\left(\frac{2\pi }{a_x}\right)^2\int\frac{\mathrm{d}^2k\mathrm{d}\omega}{(2\pi)^3}\left[(\partial_{k_ya_y}\epsilon_{\vec{k}})\partial_{k_ya_y}G(\vec{k},\omega)+(\partial_{k_ya_y}^2\epsilon_{\vec{k}})G(\vec{k},\omega)\right]\notag\\
&=&\left(\frac{2\pi }{a_x}\right)^2\int\frac{\mathrm{d}^2k\mathrm{d}\omega}{(2\pi)^3}\left[-(\partial_{k_ya_y}^2\epsilon_{\vec{k}})G(\vec{k},\omega)+(\partial_{k_ya_y}^2\epsilon_{\vec{k}})G(\vec{k},\omega)\right]=0.
\end{eqnarray}
The same derivation applies to $\Pi_{xy}(0)$, $\Pi_{yx}(0)$, and $\Pi_{yy}(0)$.

\section{8. The dominant self-energy correction of bosons}
In this section, we discuss the boson self-energy correction for a general $\vec{q}$ and $\nu$.  To the leading order in $q$, the contribution of the left diagram in Fig.~\ref{fig:boson} is given by
\begin{eqnarray}
\Pi_{ab}(\nu,\vec{q})&=&\int\frac{\mathrm{d}^dk\mathrm{d}\omega}{(2\pi)^{d+1}}v_{n\vec{k},n(\vec{k}+\vec{q})}^av_{n(\vec{k}+\vec{q}),n\vec{k}}^bG_n(\vec{k},\omega)G_n(\vec{k}+\vec{q},\omega+\nu)\notag\\
&=&\int\frac{\mathrm{d}^dk}{(2\pi)^d}v_{n\vec{k},n(\vec{k}+\vec{q})}^av_{n(\vec{k}+\vec{q}),n\vec{k}}^b\frac{f(\epsilon_{n\vec{k}})-f(\epsilon_{n(\vec{k}+\vec{q})})}{\nu+i\delta-(\epsilon_{n(\vec{k}+\vec{q})}-\epsilon_{n\vec{k}})}\notag\\
&\simeq&\int\frac{\mathrm{d}^dk}{(2\pi)^d}\delta(\epsilon_{n\vec{k}})v_{n\vec{k},n\vec{k}}^av_{n\vec{k},n\vec{k}}^b\frac{\hat{q}\cdot\vec{\nabla}_{\vec{k}}\epsilon_{n\vec{k}}}{\nu/q+i\delta-\hat{q}\cdot\vec{\nabla}_{\vec{k}}\epsilon_{n\vec{k}}}.
\end{eqnarray}
As discussed in the previous section, the constant term
\begin{eqnarray}
\Pi_{ab}(0)&=&-\int\frac{\mathrm{d}^dk}{(2\pi)^d}\delta(\epsilon_{n\vec{k}})v_{n\vec{k},n\vec{k}}^av_{n\vec{k},n\vec{k}}^b
\end{eqnarray}
is exactly cancelled by the diamagnetic term (the right diagram in Fig.~\ref{fig:boson}). The imaginary part is given by
\begin{eqnarray}
\text{Im}\Pi_{ab}(\nu,\vec{q})&=&-\pi\frac{\nu}{q}\int\frac{\mathrm{d}^dk}{(2\pi)^d}\delta(\epsilon_{n\vec{k}})v_{n\vec{k},n\vec{k}}^av_{n\vec{k},n\vec{k}}^b\delta(\nu/q-\hat{q}\cdot\vec{\nabla}_{\vec{k}}\epsilon_{n\vec{k}})\notag\\
&\simeq&-\pi\frac{\nu}{q}\int\frac{\mathrm{d}^dk}{(2\pi)^d}\delta(\epsilon_{n\vec{k}})v_{n\vec{k},n\vec{k}}^av_{n\vec{k},n\vec{k}}^b\delta(\hat{q}\cdot\vec{\nabla}_{\vec{k}}\epsilon_{n\vec{k}}).
\end{eqnarray}

\section{9. The band width of the electron band under magnetic field}
Here we show a simple numerical result on the band width of the electron band structure under a uniform magnetic field, in order to support the claim
\begin{equation}
\text{(band width)}\propto e^{-C\ell^2/a^2}.\label{band}
\end{equation}
Here $\ell=(eB)^{-1/2}$ is the magnetic length and $a$ is the lattice constant of the tight binding model.  In the continuum limit $a\rightarrow0$, the Landau levels are flat.  For a finite $a$, the lattice periodic potential produces nonzero dispersions. 

By denoting the number of the flux per a unit cell $\phi$
\begin{equation}
\frac{\ell^2}{a^2}=\frac{1}{eB a^2}=\frac{1}{\phi}
\end{equation}
Eq.~\eqref{band} suggests that 
\begin{equation}
\log{(\text{band width})}=(\text{const.}) -C\phi^{-1}.
\end{equation}

\begin{figure}[h!]
\begin{center}
\includegraphics[width=0.5\columnwidth]{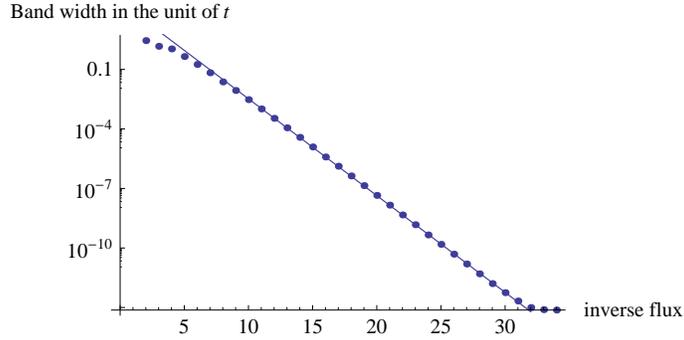}
\end{center}
\caption{The band width of the lowest landau level in the tight binding model as a function of the inverse flux $\phi^{-1}$.}
\label{fig:bandwidth}
\end{figure}

In Fig.~\ref{fig:bandwidth} we show the numerical result for the lowest Landau levels in the square lattice tight-binding model with the nearest neighbor hopping. The logarithm of the band width is indeed proportional to $\phi^{-1}$. This result holds for other Landau levels as well, as long as the van-Hove singularity energy is avoided.

\end{document}